# Combining Blockchain and IoT for Decentralized Healthcare Data Management

Sajad Meisami[1], Sadaf Meisami[2], Melina Yousefi[3] and Mohammad Reza Aref[4]

[1]Department of Computer Science, Illinois Institute of Technology, Chicago, USA
[2]Department of Management, Kharazmi University, Tehran, Iran
[3]Department of Industrial Engineering, Isfahan University of Technology, Isfahan, Iran
[4]Department of Electrical Engineering, Sharif University of Technology, Tehran, Iran

*ABSTRACT*

*The emergence of the Internet of Things (IoT) has resulted in a significant increase in research on e-health. As the amount of patient data grows, it has become increasingly challenging to protect patients' privacy. Patient data is commonly stored in the cloud, making it difficult for users to control and protect their information. Moreover, the recent rise in security and surveillance breaches in the healthcare industry has highlighted the need for a better approach to data storage and protection. Traditional models that rely on third-party control over patients' healthcare data are no longer reliable, as they have proven vulnerable to security breaches. To address these issues, blockchain technology has emerged as a promising solution. Blockchain-based protocols have the potential to provide a secure and efficient system for e-health applications that does not require trust in third-party intermediaries. The proposed protocol outlined in this paper uses a blockchain-based approach to manage patient data securely and efficiently. Unlike Bitcoin, which is primarily used for financial transactions, the protocol described here is designed specifically for e-health applications. It employs a consensus mechanism that is more suitable for resource constrained IoT devices, thereby reducing network costs and increasing efficiency. The proposed protocol also provides a privacy-preserving access control mechanism that enables patients to have more control over their healthcare data. By leveraging blockchain technology, the protocol ensures that only authorized individuals can access the patient's data, which helps prevent data breaches and other security issues. Finally, the security and privacy of the proposed protocol are analysed to ensure that it meets the necessary standards for data protection. The protocol's effectiveness and efficiency are tested under different scenarios to ensure that it can perform reliably and consistently. Finally, the protocol proposed in this paper shows that how blockchain can be used to provide a secure and efficient system that empowers patients to take control of their healthcare data.*

## 1. INTRODUCTION

The world is experiencing a significant increase in the number of patients, and the availability of primary doctors or medical staff is becoming more challenging. This situation has led to a growing interest in using the Internet of Things (IoT) in healthcare to improve medical services. The IoT allows any device to connect to other devices and the internet at any time and place, making it a powerful tool in addressing the challenges in the healthcare programmes. According to research, over 75 billion devices will be able to connect to the internet by 2025, presenting vast opportunities for IoT applications in e-health [1]. E-health is one of the main applications of IoT, and it offers many benefits for both medical staff and patients. One significant advantage of e-health is that it allows medical staff to treat more patients while providing more convenience for patients. With e-health, patients can stay connected with their doctors or medical staff as required, reducing the need for frequent visits to healthcare facilities. This, in turn, reduces medical costs and improves the quality of care and treatment. Additionally, e-health can help medical staff to monitor patient health more effectively, leading to early detection of health issues and timely intervention.

The advent of the Internet of Things (IoT) has revolutionized the healthcare programmes, and wearable devices have emerged as a significant contributor to the development of e-health. Wearable devices like smartwatches and fitness trackers can measure a patient's vital signs, including heart rate, blood glucose, body temperature, and blood pressure, among others. This capability has opened a vast array of opportunities for remote patient monitoring and disease management. One of the most significant benefits of wearable devices is that they collect patient health data automatically, eliminating the need for manual data collection by healthcare providers. This data is then transmitted to central storage or the cloud, where it can be analysed and processed by medical staff to facilitate health monitoring, disease diagnosis, and treatment. The use of wearable devices for health monitoring has made it easier for medical staff to provide personalized care to patients, allowing them to monitor the patient's health status in real-time and make timely interventions when necessary. Furthermore, the collected data can be used to detect patterns and trends in a patient's health status over time. This information can be analyzed to predict future health problems, enabling medical staff to provide proactive care and treatment. Wearable devices also allow patients to monitor their own health, encouraging them to take an active role in managing their health and well-being [2].

The protection of patients' healthcare data is of utmost importance as it contains sensitive and confidential information that must be kept secure. However, with the increasing use of technology in the healthcare application, storing and processing patients' data remotely has become a common practice. This has led to concerns about the privacy and confidentiality of patients' healthcare data, as the data is vulnerable to unauthorized access and malicious attacks. One of the most significant risks associated with remote data storage and processing is the potential for security breaches. Malicious actors can exploit vulnerabilities in the system to access patients' data, and once they gain access, they can modify or replace the data with incorrect information or steal it entirely. This type of security attack can have severe consequences for patients, leading to misdiagnosis, mistreatment, and potential harm [3].

A. **Organization**

Section II describes the related work involved. Section III discusses the challenges we solve in this paper. Sections IV provides an overview of the system model, whereas section V explains the network protocol in detail. Section VI discusses the security and privacy of the proposed model, and section VII provides a conclusion.

## 2. RELATED WORKS

To address the privacy issue on personal data, researchers developed various methods. One of them is data anonymization that attempts to protect personally identifiable information. For example, in the k-anonymity method used in anonymous datasets, any necessary recorded data is indistinguishable from at least k−1 other important recorded data [4]. However, in recent research, it has been shown, anonymized datasets can be broken even with a little information (their anonymity disappears) [5]. There exist other privacy-preserving methods like differential privacy, that perturbs data, or adds noise to the computational process before sharing the data [6] and practices on creating noisy data or summarizing [7]. These methods are not efficient in healthcare applications where patients' original data are required to send to medical staff for medical treatments.

Attribute-based encryption (ABE) is a useful technology that can provide data privacy and fine-grained access control when users want to secretly share data stored in a third-party cloud server [8]. Almost all ABE schemes require a trusted private key generator (PKG) to set up the system and distribute for users the corresponding secret key [9]. However, in all ABE schemes, the PKG can decrypt all data stored in the cloud server, which may cause serious problems such as privacy data leakage and key abuse. Furthermore, the traditional cloud storage model runs in a centralized storage manner, so existence a single point of failure can collapse the system. Other encryption schemes exist that allow running computations and queries over encrypted data that called fully homomorphic encryption (FHE) [10] but are currently too unsuitable to use in practice widely.

In recent years, decentralized cryptocurrency systems have emerged. Bitcoin was the first of these systems, which use blockchain technology. Bitcoin allows users securely to make transactions and transfer currency (bitcoins) with others without the need to trusted third-party [11]. Blockchain works as an immutable timestamp ledger of blocks that are shared across all participating nodes in the network, which can bypass the need for a central authority [12]. This technology is used for sharing and storing data in a distributed manner by a peer-to-peer network [13]. Nowadays, blockchain is playing an effective role in financial transactions [14]. Also, it can be a facilitator in many other fields. Such as decentralized IoT [15], identity-based PKI [16], decentralized supply chain [17], decentralized proof of document existence [18], decentralized storage [19–21], etc.

In [22], the work by Zyskind et al. has shown the use of blockchain technology to construct an access control and management platform for personal data. They focused on users' privacy. And they combine blockchain and off-chain storage to storing encrypted data out of blockchain ledger while pointers to the encrypted data exist on the blockchain. Recently researchers addressed the security and privacy issues on healthcare data, using blockchain technology, and proposed new schemes [23–26].

In this paper, we combine IoT and blockchain technology to construct a novel platform for patients' healthcare data management that satisfies privacy and security requirements.

## 3. CHALLENGES AND OUR SOLUTIONS

The primary focus of e-health applications in the IoT is to ensure the secure transmission and preservation of patients' healthcare data. The decentralized nature of blockchain technology, along with its attributes such as immutability and transparency, make it a promising solution for achieving these goals. However, despite its potential benefits, there are still challenges associated with the implementation of blockchain technology in the IoT. These challenges may include resource constraints, scalability issues, and high computational requirements. In this context, it is crucial to develop effective solutions to address these challenges and enable the successful implementation of blockchain technology in e-health IoT applications.

### 3.1. Scalability

Due to their limited resources, IoT devices face challenges in executing computationally intensive tasks, such as solving problems for adding new blocks to the blockchain ledger through consensus algorithms. Proof-of-Work (PoW) is a commonly used consensus algorithm, but its high computational requirements make it unsuitable for resource-constrained IoT devices. To address this challenge, we have adopted an alternative consensus method called Practical Byzantine Fault Tolerance (PBFT) for our blockchain network [27]. PBFT is a voting-based consensus method that involves multiple rounds of voting by all nodes in the network [28]. By eliminating the need for PoW, we can significantly reduce the network costs associated with consensus operations, including bandwidth, processor requirements, and energy consumption. Overall, using PBFT as a consensus method allows us to increase the network's efficiency while minimizing the resource demands on IoT devices. This approach can help ensure the smooth operation of e-health applications on IoT devices, which rely on the efficient processing and transmission of sensitive patient data.

## 3.2. Data Storage

Storing IoT big data directly on the blockchain ledger is not a practical or suitable approach due to a large amount of storage space required. To address this challenge, we have developed a solution that involves storing only pointers to the data (hashed encrypted data) on the blockchain while storing the actual data (encrypted form) on off-chain storage, as detailed in section IV, E. Off-chain Storage. By using this approach, we can significantly reduce the storage requirements of the blockchain while still ensuring that the data is securely stored and accessible. The use of off-chain storage for IoT big data also helps to minimize the processing and bandwidth demands on IoT devices, enabling them to operate more efficiently. Overall, this approach to storing IoT big data strikes a balance between the need for secure and accessible data and the practical limitations of blockchain technology. It represents an innovative solution to a significant challenge in the development of e-health applications on IoT devices.

## 3.3. Security of Data

To ensure the security of patients' healthcare data, we employ a symmetric key encryption scheme, which is a cryptographic technique that provides confidentiality and integrity to data. In this scheme, the data is encrypted using a symmetric key, and only authorized parties can access this key for decryption. We use this encryption technique to protect the data during transmission and storage. As a result, even when the data is stored off-chain, it remains encrypted, and only authorized parties with the key can access and decrypt the data. This approach provides an additional layer of protection to sensitive healthcare data, ensuring that only authorized parties can access it.

## 3.4. Patients' Privacy

The main concern that is addressed in this paper is preserving patients' privacy because Patients' healthcare data is highly privacy-sensitive. We assume that the medical staff is honest-but-curious (i.e., they follow the protocol). In our system, patients could remain (pseudo) anonymous. At the same time, medical staff profiles could be stored on the blockchain so that patients can trust medical staff by verifying medical staff identities. Our proposed platform satisfies the following Items:

1) Patients' Data Ownership: Our proposed platform upholds the concept of patients' data ownership, where patients have the sole authority to control their healthcare data. In this system, we acknowledge medical staff as Service Providers who are granted specific permissions set by the patients. Therefore, patients hold the position of healthcare data owners and have complete control over their data. This approach enables patients to maintain their privacy and have the freedom to share their healthcare data with selected medical staff, thereby ensuring a high level of trust between patients and medical staff.

International Journal on Cryptography and Information Security (IJCIS), Vol. 13, No.1, March 2023

2) Fine-grained Access Control: Each patient can grant a set of permissions to Any desired member of the medical staff for accessing a patient's healthcare data. Also, the patient can alter or revoke the set of granted access permissions. These permissions are securely stored on blockchain ledger as access-control policies, where only the patient can change or revoke them.
3) Data Transparency and Auditability: Patients have complete and accurate transparency over their collected healthcare data, and they can know how medical staff can access to which part or type of their healthcare data.

## 4. OUR SYSTEM MODEL

As demonstrated in Figure 1, our proposed system model consists of the following main modules:

1) Wearable IoT devices.
2) Patient's smartphone.
3) Medical staff.
4) Blockchain.
5) Off-chain storage.

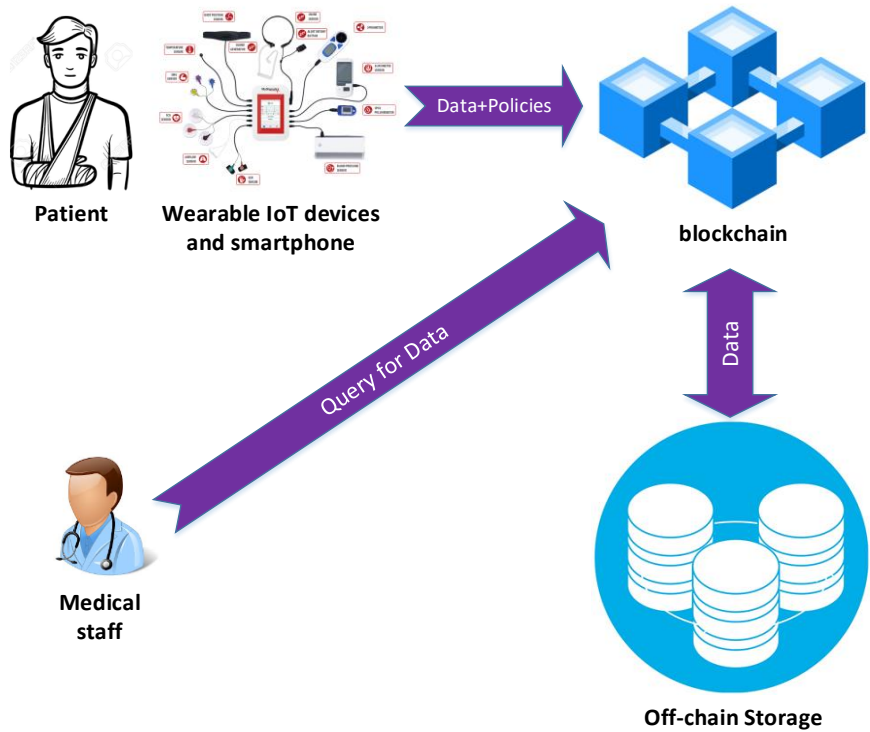

Fig. 1. Our proposed system model.

### 4.1. Wearable IoT devices

The healthcare data collected by wearable IoT devices include vital signs such as blood pressure, body temperature, and heart rate. However, these devices have resource limitations, such as low energy, processing power, and storage space. To address these limitations, the collected data is sent to the patient's smartphone through short-range communication methods such as Bluetooth or Zigbee. This allows wearable devices to conserve their resources while still collecting the necessary data, which can then be transmitted to the blockchain network for further analysis.

### 4.2. Patient's Smartphone

The computing capabilities of smartphones are greater than those of IoT devices, as smartphones have more storage space, longer battery life, and higher processing power. Therefore, smartphones can perform more complex tasks, such as computational and cryptographic operations. Additionally, smartphones can transmit data using long-range communication methods, such as cellular networks, and act as a gateway, enabling patients to connect with the blockchain network.

### 4.3. Medical Staff

Medical staff, including physicians and nurses, should receive patients' healthcare data and, after analyzing it, obtain information about the patient's health status. Then they provide appropriate treatment for patients.

### 4.4. Blockchain

In simple terms, blockchain is a decentralized and secure digital ledger that maintains a record of transactions. It is a chain of blocks, where each block contains a set of transactions and a cryptographic hash of the previous block. Blockchain technology is designed to be tamper-proof, making it highly secure and resistant to any unauthorized modifications. Because of its transparency and decentralized nature, blockchain is often used to facilitate secure transactions between parties without the need for intermediaries.

In our work, we use blockchain to store access policies and eliminate the need for a third-party that preserves network against DoS attack and single point of failure. It also ensures the availability and integrity of the patients' data. The data is not stored on the blockchain, but only the pointers to the data (hash of encrypted data) are stored on it to lighten the storage space of blockchain. Also, because the use of PoW in IoT applications is not appropriate, we use PBFT for consensus operations.

### 4.5. Off-Chain Storage

We store patients' encrypted data on off-chain storage. For the implementation of off-chain storage, we use The InterPlanetary File System (IPFS) [21] that is a peer-to-peer distributed file system that seeks to connect all computing devices with the same system of files. It provides a high throughput content-addressed block storage model, with content-addressed hyperlinks. IPFS combines a distributed hashtable (or DHT), an incentivized block exchange, and a self-certifying namespace. IPFS has no single point of failure, and nodes do not need to trust each other [21]. In IPFS, we distribute the data and store them on different servers all over the world. Not using the central server is the reason for the advantage of IPFS over conventional cloud storage.

## 5. THE NETWORK PROTOCOL

In this section, we describe the our system protocol design to ensure the privacy and security of patients' healthcare data while allowing medical staff to access and analyze it for the purpose of providing appropriate treatment.

### 5.1. Cryptographic Techniques used in the System

*Hash Function*: We use SHA-256 to implement the hash function (H indicates the hash function). SHA-256 (Secure Hash Algorithm 256-bit) is a cryptographic hash function that takes input data of variable length and produces a fixed-length output of 256 bits. It is one of the most widely used cryptographic hash functions and is commonly used to ensure data integrity and message authentication [29]. In order to perform its cryptographic functions, SHA-256 breaks down input data into blocks of a fixed size and processes each block through a series of mathematical operations. The output of each block's processing is combined with the result of the previous block until the entire message has been processed. Regardless of the input data's size, the output of SHA-256 is a fixed-length string of 64 hexadecimal characters that is unique to the input data. To ensure data integrity and prevent tampering, SHA-256 provides a secure hashing function that cannot be reversed to obtain the original input data from the output hash. This feature makes it a suitable choice for various security applications, including digital signatures algorithms and data integrity checks.

*Symmetric Key Encryption*: Symmetric algorithm uses the same key for encryption of plaintext and decryption of ciphertext. We use AES to implement the symmetric key encryption. AES (Advanced Encryption Standard) [30] is a popular symmetric block cipher encryption algorithm that was introduced by the National Institute of Standards and Technology (NIST) in 2001 to replace the less secure Data Encryption Standard (DES). AES is widely used to secure sensitive data in various applications, including electronic communication, and banking applications. To encrypt and decrypt data, AES employs a symmetric key, which means that the same key is used for both encryption and decryption. The key length, which can be either 128, 192, or 256 bits, determines the strength of the encryption. AES operates by dividing the plaintext into fixed-sized blocks and then performing a series of mathematical operations on each block known as rounds. The number of rounds required is determined by the key length, with 10 rounds for 128-bit keys, 12 rounds for 192-bit keys, and 14 rounds for 256-bit keys. The security of AES is based on the fact that it is computationally infeasible to determine the private key which is used for encryption from the ciphertext without knowledge of the key. This fact makes AES a strong encryption algorithm that perfectly functions in real-world applications and is widely used to secure sensitive information. ($G_{enc}$ indicates generating algorithm).

*Digital Signature Scheme*: A digital signature is a cryptographic technique used to verify the authenticity and integrity of a digital document or message. It involves generating a unique digital signature using a mathematical algorithm, which can only be created by the owner of a private key in a public key infrastructure (PKI). The process of creating a digital signature involves first generating a hash of the message using a hashing algorithm like SHA-256. Afterward, the sender encrypts the hash value using their private key to produce a unique digital signature. This digital signature is then appended to the message together with the sender's public key, which can be used to authenticate the signature's origin and verify the integrity of the message. After receiving the message with the digital signature and the sender's public key, the recipient can use the public key to decrypt the signature and obtain the hash value of the message. The recipient then generates a hash of the received message using the same hashing algorithm and compares it with the hash value obtained from the digital signature. If the two hash values match, the recipient can be assured that the message has not been altered during transmission and that the signature was created by the sender with their private key, which confirms the sender's authenticity. ECDSA stands for Elliptic Curve Digital Signature Algorithm. It is a type of digital signature algorithm that uses elliptic curve cryptography to provide authentication and integrity for digital documents or messages. ECDSA is widely used in applications such as digital identity verification, financial transactions, and secure communication, due to its strong security properties and efficient use of resources. In our protocol Digital signature is added to the data for authentication purposes. For the implementation of the digital signature, we use ECDSA with a secp256k1 curve [31] ($G_{sig}$ indicates generating algorithm).

## 5.2. Protocol in Detail

staff *m* each generates a pair of private and public keys to sign and send transactions to the blockchain network and announce their public key (as their address) on the network. Patient *P* also generates a secret key for encrypting data with an AES symmetric encryption algorithm. Then the patient shares the secret key with the chosen member of the medical staff so that, later, she will be able to decrypt her authorized data with that secret key.

---

**Protocol 1** Joining the Blockchain

---
1: **procedure** GENERATING($p,m$)
2:       $p$ executes:
3:           $(pk^p_{sig}, sk^p_{sig}) \leftarrow G_{sig}(\ )$
4:           $sk^{p,m}_{enc} \leftarrow G_{enc}(\ )$
5:       $p$ shares $pk^p_{sig}$ *(as address) on the network*
6:       $m$ executes:
7:           $(pk^m_{sig}, sk^m_{sig}) \leftarrow G_{sig}(\ )$
8:       $m$ shares $pk^m_{sig}$ *(as address) on the network*
9:       $p$ shares $sk^{p,m}_{enc}$ *with m from secure channel*
10:      // Both $p$ and $m$ have $sk^{p,m}_{enc}$
11:      **return** $pk^p_{sig}, pk^m_{sig}, sk^{p,m}_{enc}$
12: **end procedure**

---

*Registration of access policy:* We denote the data access permissions by $POLICY_{p,m}$, which indicates the permissions that the patient *p* gives to the selected member of the medical staff *m* so that she can access a particular type or all of the patient's data. For example, $POLICY_{p,m} = \{body\ temperature,\ blood\ pressure\}$.

---

**Protocol 2** Access Control Transaction

---
1: **procedure** ACCESSTX($pk^k_{sig}$, $message$)
2:       $s \leftarrow 0$
3:       $(pk^p_{sig}\ ||\ pk^m_{sig}\ ||\ POLICY_{p,m}) = message$
4:       **if** $pk^k_{sig} = pk^p_{sig}$ **then**
5:           $L[H(pk^k_{sig})] \leftarrow L[H(pk^k_{sig})] \cup message$
6:           // $L$ is Blockchain memory
7:           $s \leftarrow 1$
8:      **end if**
9:      **return** $s$
10: **end procedure**

---

By sending a $T_{access}$ transaction on the blockchain network that contains $POLICY_{p,m}$, the patient gives the desired permissions to the medical staff. As illustrated in protocol 2, this transaction is performed in the nodes of the blockchain, and it is checked the patient himself has sent the transaction, then the set of permissions are stored on the blockchain ledger.

The patient can send new $T_{access}$ transactions and change the permissions set granted to the medical staff. Also, Sending the empty set by the patient can revoke all access-permissions set previously granted.

*Data storage and retrieval:* $T_{data}$ transaction is used to store patients' encrypted healthcare data on off-chain storage (IPFS) or access stored data and receive it. The patient (to store and retrieve the data) and the medical staff (only to retrieve the data) can send the $T_{data}$ transaction to the blockchain network.

If the $T_{data}$ transaction (by the patient or the medical staff) is sent to the network, the nodes in the blockchain first check with the following protocol (protocol3) whether they have access permissions or not?

**Protocol 3** Blockchain Permissions Checking
1: **procedure** POLICYCHECK($pk^k_{sig}$, T)    //T=type of data
2:     $s \leftarrow 0$
3:     **if** $L[H(pk^k_{sig})] \neq \emptyset$ **then**
4:         $(pk^p_{sig} || pk^m_{sig} || POLICY_{p,m}) \leftarrow L[H(pk^k_{sig})]$
5:         **if** $pk^k_{sig}=pk^p_{sig}$ **or**
6:             ($pk^k_{sig}=pk^m_{sig}$ and $T \in POLICY_{p,m}$) **then**
7:                 $s \leftarrow 1$
8:         **end if**
9:     **end if**
10:    **return** $s$
11: **end procedure**

Now, after checking the access permissions and the transaction sender's approval, he or she can store or retrieve patients' encrypted healthcare data with the following protocol.

**Protocol 4** Data Transaction
1: **procedure** DATATX($pk^k_{sig}$, message)
2:     (C || T || RW) = message
3:     // C=encrypted data , T=type of data
4:     //RW=read data(=1) or write data(=0)
5:     **if** POLICYCHECK($pk^k_{sig}$, T)=True **then**
6:         $(pk^p_{sig} || pk^m_{sig} || POLICY_{p,m}) \leftarrow L[H(pk^k_{sig})]$
7:         **if** RW = 0 **then**
8:             $L[pk^k_{sig} || T] \leftarrow L[pk^k_{sig} || T] \cup H(C)$
9:             (IPFS) $ds[H(C)] \leftarrow C$
10:            **return** $H(C)$
11:        **else if** $C \in L[pk^k_{sig} || T]$ **then**
12:            (IPFS) **return** $ds[H(C)]$
13:        **end if**
14:    **end if**
15:    **return** $\emptyset$
16: **end procedure**

Note that we used *IPFS* shorthand notation in lines 9 and 12 of Protocol 4 for accessing the off-chain storage. *IPFS* instruction cause to send Off-blockchain network message in off-chain storage for storing or retrieving data.

With the above protocols, the patient can easily upload the encrypted healthcare data in the network. The chosen members of the medical staff can also receive the encrypted data if there is a permission, and then with the $sk^{p,m}_{enc}$ that they have received before (in protocol 1), they can decrypt encrypted data and access the original healthcare data.

## 6. SECURITY AND PRIVACY ANALYSIS

In this section, we discuss and investigate the performance of our proposed protocol in terms of security and privacy. For security designing in any model, there are exist three main security requirements that need to be addressed: Confidentiality, Integrity, and Availability, known as CIA [32]. Confidentiality makes sure that the system's messages should be read by only authorized users who can access the system. The data integrity is responsible for making sure that no one without permission can change the stored data, and the availability of the data means that when users needed to the data, it is available to them. Now, we summarize the aforementioned primary security requirement evaluation in Table 1.

Table 1. Security Requirement Analysis.

| Requirement | Model Solution |
| --- | --- |
| Confidentiality | Achieved by using symmetric key encryption. |
| Integrity | Hashing of data blocks in blockchain is employed to achieve integrity. |
| Availability | Achieved by limiting acceptable transactions in the network. |
| Authorization | Using digital signature to achieve authorization. |

In our model, data ownership is emphasized for preserving privacy. That means only the patients (users) can control their data. The decentralized nature of the blockchain technology and using digital signature to sign transactions in the network ensure that an adversary cannot be able to infiltrate the system as a user. Because gaining control over the majority of the network's resources (at least 51%) or forging digital-signature is almost impossible for the adversary. Also, the model ensures other privacy-preserving parameters that we previously mentioned (in section III) like, Data Transparency and Auditability. Fine-grained access control is satisfied by storing access-control policies on a blockchain ledger, where only the patient can change or revoke them. In Table 2, we have made a comparison between our proposed model and other existing systems and show how are work outperformed the other existing works. One of the important points is that our proposed model is using decentralized and distributed off-chain storage space to prevent single point of failure.

Table 2. Comparison of Different Existing Systems.

| Model Name / Property | Yang[33] | Xia-I[34] | Xia-II[36] | Peterson[38] | Zang[40] | A.Zhang[42] | Our Proposed Model |
| --- | --- | --- | --- | --- | --- | --- | --- |
| Access control | ✓ | ✓ | ✓ | ✓ | ✓ | ✓ | ✓ |
| Blockchain-Based | ✗ | ✓ | ✓ | ✓ | ✗ | ✓ | ✓ |
| Privacy Preserving | ✓ | ✓ | ✓ | ✓ | ✓ | ✓ | ✓ |
| IPFS Off-chain storage | ✗ | ✗ | ✗ | ✗ | ✗ | ✗ | ✓ |

Now, we describe a few significant real-world attacks and analyze the resilience of our model against each of them.

### 6.1. Denial of Service (DOS) Attack

Generally, a DoS (Denial of Service) attack is a malicious cyber-attack that disrupts the regular operation of a system or network by inundating it with an excessive amount of traffic or requests. The objective is to render the targeted system or network inaccessible to legitimate users by using up its resources, such as bandwidth or processing power. This can result in a significant impact on the target's business, finances, or reputation. Attackers can use different tactics, such as flooding the target with traffic, exploiting vulnerabilities in the system, or leveraging malware to control a network of compromised devices to carry out the attack. Specially, in our case of study an attacker generates a large number of transactions to increase traffic in the network and disrupt the blockchain.

Defence: Only two types of transactions can be sent in the network. Also, each node can send a limited number of transactions, and the blockchain network will reject the rest of the user's transactions after receiving a few messages from a specific node. So, our proposed model has high resilience against DoS attack.

### 6.2. Modification Attack

A modification attack is a form of cyber-attack where the attacker tries to modify data or information in transit between two parties, often without being detected. The attack can happen at any stage in the communication process, from transmission over a network to storage on a device. The objective of a modification attack is to modify the data in a manner that benefits the attacker, such as changing the message's contents, redirecting the destination of a transfer, or manipulating the values of financial transactions. To prevent modification attacks, techniques such as encryption and digital signatures can be employed to ensure the data's integrity and authenticity during transmission. Specially, in our case of study an attacker tries to modify or remove the stored patients' data (like access policies and hash) on the blockchain ledger.

Defence: Blockchain, by using an immutable ledger and other techniques such as encryption and digital signatures can ensure the data's integrity and authenticity during transmission and create high resistance against modification attack.

### 6.3. Public Blockchain Modification

A public blockchain can also be vulnerable to modification attacks, where an attacker tries to alter the contents of a block in the chain, resulting in the creation of a new invalid chain or a fork in the existing chain. Public blockchains are decentralized and transparent ledgers that allow anyone to participate in the network.

Defence: To prevent modification attacks, a consensus mechanism is used to ensure that all nodes in the network agree on the state of the chain, and any modifications require validation by the majority of nodes in the network. We used PBFT consensus method in our protocol. Also, we can use the private type of blockchain, so the nodes are from outside the organization cannot work as miners to create a malicious block.

### 6.4. Storage Attack

Generally, A storage attack is a cyber-attack that involves unauthorized access to data that is stored on a system or device. These attacks can be carried out using various methods, such as exploiting security vulnerabilities, using malware to access sensitive information, or physically accessing the storage medium. The aim of a storage attack is to steal or manipulate the data that is being stored, which could include financial information, personal data, or other valuable assets. In our case of study, an attacker wants to remove, change, or add data in the Off-chain storage.

Defence: To prevent storage attacks, security measures such as encryption, access control, and regular data backups can be implemented. In our system, on blockchain ledger exist a hash of the encrypted data stored in the Off-chain storage; therefore, changes in the data can easily be detected. So, there is a high resilience against this attack.

### 6.5. Appending Attack

An appending attack is a cybersecurity attack where an unauthorized user gains access to a system or network and adds or appends data or commands to a file or program. The objective of such an attack is to execute malicious code, steal sensitive data or take over the targeted system. The attacker can append their own code to a program to manipulate the program's behaviour or add commands to an existing script, thereby granting them unauthorized access to the system. Specifically, in our study an attacker tries to compromise a miner and generate blocks with fake transactions to create a false reputation.

Defence: Due to the usage of private blockchain, so the users cannot generate a fake block, whereas miners in the blockchain must verify any transaction.

### 6.6. 51% Attack

In general, an attack known as 51% attack can be executed on blockchain networks, particularly those using proof-of-work (PoW) consensus algorithms. In this attack, an attacker acquires control of more than 50% of the network's computing power or hash rate, enabling them to manipulate the blockchain ledger. As a result, they can double-spend cryptocurrency, prevent new transactions from being confirmed, or even reverse previously confirmed transactions. The name "51% attack" is derived from the fact that the attacker must have at least 51% of the network's computing power to carry out the attack successfully.

Defence: To prevent 51% attacks in our blockchain network, we use alternative consensus algorithms named PBFT which create high resilience against this attack.

### 6.7. Distributed DOS (DDOS) Attack

This is a distributed version of the Denial of Service (DOS) Attack.

Defence: A valid node can send a limited number of transactions in the network. As a defence, when the blockchain network receives a transaction, miners check that received transaction has produced by a valid node then accept it.

Table 3.  Security Analysis Against Attacks

| Attack | Definition | Defence | Resilience |
|---|---|---|---|
| Denial of Service (DOS) Attack | Attacker generates a large number of transactions to increase traffic in the network and disrupt the blockchain. | Only two types of transactions can be sent in the network. Also, each node can send a limited number of transactions, and the blockchain network will reject the rest of the user's transactions after receiving a few messages from a specific node. | High |
| Modification Attack | Attacker modifies or removes the stored patients' data (like access policies and hash) on the blockchain ledger. | Blockchain uses an immutable ledger. | High |
| Public blockchain Modification | Attacker advertises a false ledger of blocks and makes it as the longest ledger. | We used PBFT consensus method to prevent it. Also, we can use the private type of blockchain, so the nodes are from outside the organization cannot work as miners to create a malicious block. | High |
| Storage Attack | Attacker wants to remove, change, or add data in the Off-chain storage. | On blockchain ledger exist a hash of the encrypted data stored in the Off-chain storage; therefore, changes in the data can easily be detected. | High |
| Appending Attack | Attacker compromises a miner and generates blocks with fake transactions to create a false reputation. | Due to the usage of private blockchain, so the users cannot generate a fake block, whereas miners in the blockchain must verify any transaction. | High |
| 51% Attack | Attacker controls more than 51% of miners and tries to compromise the consensus algorithm and generate a fake block. | The probability of occurrence of this attack is very low due to the usage of private blockchain and PBFT method for consensus. | High |
| Distributed DOS (DDOS) Attack | This is a distributed version of the Denial of Service (DOS) Attack. | A valid node can send a limited number of transactions in the network. After the blockchain network receives a transaction, miners check that received transaction has produced by a valid node then accept it. | Moderate |

## 7. CONCLUSION

Patients' healthcare data are privacy-sensitive and security-sensitive, and for managing them, we should not trust the third-parties, where they are vulnerable to attacks and abuse. In this work, we proposed a novel platform based on the Internet of Things (IoT) and blockchain technology, motivated by the privacy and security challenges of patients' healthcare data in e-health. Our proposed platform enables patients to have ownership and full control over their sensitive healthcare data collected by their IoT wearable devices. This ownership and complete control over patients' data are satisfied by storing access control policies in a blockchain ledger by patients to specify who from medical staff can access patients' data. In this platform, we use off-blockchain storage to lighten the blockchain storage. Also, we use a suitable consensus method in the blockchain network due to the resource constraint factor of IoT.

## ACKNOWLEDGEMENT

The authors are thankful to Mr. Mohammad Doost and Mr. Mohammadhadi Ahmadiashtiyani for consulting and lengthy discussions on many of the ideas used in this paper.

## REFERENCES

[1] Statista, "Internet of Things (IoT) connected devices installed base worldwide from 2015 to 2025. [Online]. Available: https://www.statista.com/statistics/471264/iot-number-of connected-devices-worldwide/

[2] S. M. R. Islam, D. Kwak, M. H. Kabir, M. Hossain and K.-S. Kwak, "The Internet of Things for Health Care: A Comprehensive Survey," IEEE Access, vol. 3, pp. 678 - 708, June 2015.

[3] K. Abouelmehdi, A. Beni-Hessane and H. Khaloufi, "Big healthcare data preserving security and privacy," Journal of Big Data, December 2018.

[4] Latanya Sweeney. k-anonymity: A model for protecting privacy. International Journal of Uncertainty, Fuzziness and Knowledge-Based Systems, 10(05):557–570, 2002.

[5] Arvind Narayanan and Vitaly Shmatikov. How to break anonymity of the netflix prize dataset. arXiv preprint cs/0610105, 2006.

[6] Cynthia Dwork. Differential privacy. In Automata, languages and programming, pages 1–12. Springer, 2006.

[7] Y.-A. d. Montjoye, E. Shmueli, S. S. Wang and A. S. Pentland "Protecting the Privacy of Metadata through SafeAnswers," PLOS ONE vol. 9, July 2014.

[8] A.Sahai and B.Waters, ''Fuzzy identity-based encryption,'' in Proc.Annu. Int. Conf. Theory Appl. Cryptograph. Techn. Berlin, Germany: Springer, 2005, pp. 457–473.

[9] J. Zhang, X. A. Wang, and J. Ma, ''Data owner based attribute based encryption,'' in Proc. Int. Conf. Intell. Netw. Collaborative Syst. (INCOS), Sep. 2015, pp. 144–148.

[10] Craig Gentry. Fully homomorphic encryption using ideal lattices. In STOC, volume 9, pages 169–178, 2009.

[11] Satoshi Nakamoto. Bitcoin: A peer-to-peer electronic cash system. Consulted, 1(2012):28, 2008.

[12] A. Dorri, S. S. Kanhere, R. Jurdak and P. Gauravaram, "LSB: A Lightweight Scalable BlockChain for IoT Security and Privacy, arXiv:1712.02969 , Dec 2017.

[13] A. Kosba, A. Miller, E. Shi, Z. Wen and C. Papamanthou, "Hawk: The Blockchain Model of Cryptography and Privacy-Preserving Smar Contracts," in 2016 IEEE Symposium on Security and Privacy (SP), San Jose, CA, USA, August 2016.

[14] Blockchain for Financial Services. Accessed: Mar. 25, 2018. [Online]. Available: https://www.ibm.com/blockchain/financial-services

[15] A Decentralized Network for Internet of Things. Accessed: Mar. 25, 2018. [Online]. Available: https://iotex.io

[16] C. Fromknecht and D. Velicanu. (2014). A Decentralized Public Key Infrastructure With Identity Retention. [Online]. Available: https://eprint.iacr.org/2014/803.pdf

[17] Blockchain for Supply Chain. Accessed: Mar. 25, 2018. [Online]. Available: https://www.ibm.com/blockchain/supply-chain

[18] Proof of Existence. Accessed: Mar. 25, 2018. [Online]. Available: https://proofofexistence.com

[19] S. Wilkinson, T. Boshevski, J. Brandoff, and V. Buterin, ''Storj a peerto-peer cloud storage network,'' White Paper. Accessed: Mar. 25, 2018. [Online]. Available: https://storj.io/storj.pdf
[20] P. Labs. (2018). Filecoin: A Decentralized Storage Network. [Online]. Available: https://filecoin.io/filecoin.pdf
[21] J. Benet. (2014). ''IPFS-content addressed, versioned, P2P file system.'' [Online]. Available: https://arxiv.org/abs/1407.3561
[22] Zyskind, Guy, and Oz Nathan. "Decentralizing privacy: Using blockchain to protect personal data." In 2015 IEEE Security and Privacy Workshops, pp. 180-184. IEEE, 2015.
[23] T.-T. Kuo and L. Ohno-Machado, "ModelChain: Decentralized Privacy Preserving Healthcare Predictive Modeling Framework on Private Blockchain Networks," arXiv:1802.01746, Feb 2018.
[24] Meisami, Sajad, Mohammad Beheshti-Atashgah, and Mohammad Reza Aref. "Using Blockchain to Achieve Decentralized Privacy In IoT Healthcare." Cryptology ePrint Archive (2021).
[25] X. Yue, H. Wang, D. Jin, M. Li and W. Jiang, "Healthcare Data Gateways: Found Healthcare Intelligence on Blockchain with Nove Privacy Risk Control," Journal of Medical Systems, 2016.
[26] Dwivedi, Ashutosh Dhar, Gautam Srivastava, Shalini Dhar, and Rajani Singh. "A decentralized privacy-preserving healthcare blockchain for IoT." Sensors 19, no. 2 (2019): 326.
[27] M. Castro, B. Liskov et al., "Practical byzantine fault tolerance," in OSDI, vol. 99, 1999, pp. 173–186.
[28] M. S. Ali, M. Vecchio, M. Pincheira, K. Dolui, F. Antonelli and a. M. H Rehmani, "Applications of Blockchains in the Internet of Things: A Comprehensive Survey," IEEE Communications Society, 2018.
[29] Federal Information and Processing Standards. FIPS PUB 180-4 Secure Hash Standard ( SHS ). (March), 2012.
[30] Daemen, Joan, and Vincent Rijmen. The design of Rijndael. Vol. 2. New York: Springer-verlag, 2002.
[31] Don Johnson, Alfred Menezes, and Scott Vanstone. The elliptic curve digital signature algorithm (ecdsa). International Journal of Information Security, 1(1):36–63, 2001.
[32] N. Komninos, E. Philippou, and A. Pitsillides, "Survey in smart grid and smart home security: Issues, challenges and countermeasures," IEEE Communications Surveys & Tutorials, vol. 16, no. 4, pp. 1933–1954, 2014.
[33] Y. Yang and M. Ma, "Conjunctive keyword search with designated tester and timing enabled proxy re-encryption function for e-health clouds," IEEE Transactions on Information Forensics and Security, vol. 11, no. 4, pp. 746–759, 2015.
[34] Q. Xia, E. Sifah, A. Smahi, S. Amofa, and X. Zhang, "Bbds:Blockchain-baseddatasharingforelectronic medical records in cloud environments," Information, vol. 8, no. 2, p. 44, 2017.
[35] Doost, M., Alireza Kavousi, J. Mohajeri and M. Salmasizadeh. "Analysis and Improvement of an E-voting System Based on Blockchain." 2020 28th Iranian Conference on Electrical Engineering (ICEE) (2020): 1-4.
[36] Q. Xia, E. B. Sifah, K. O. Asamoah, J. Gao, X. Du, and M. Guizani, "Medshare: Trust-less medical data sharing among cloud service providers via blockchain," IEEE Access, vol. 5, pp. 14757–14767, 2017.
[37] Khalesi, Ali, Mahtab Mirmohseni, and Mohammad Ali Maddah-Ali. "The Capacity Region of Distributed Multi-User Secret Sharing." arXiv preprint arXiv:2103.01568 (2021).
[38] K. Peterson, R. Deeduvanu, P. Kanjamala, and K. Boles, "A blockchain-based approach to health information exchange networks," in Proc. NIST Workshop Blockchain Healthcare, vol. 1, 2016, pp. 1–10.
[39] Meisami, Sajad & Beheshti-Atashgah, Mohammad & Aref, Mohammad. (2023). Using Blockchain to Achieve Decentralized Privacy in IoT Healthcare. International Journal on Cybernetics & Informatics. 12. 97-108. 10.5121/ijci.2023.120208.
[40] J. Zhang, N. Xue, and X. Huang, "A secure system for pervasive social network-based healthcare," IEEE Access, vol. 4, pp. 9239–9250, 2016.
[41] Bodell III, William E., Sajad Meisami, and Yue Duan. "Proxy Hunting: Understanding and Characterizing Proxy-based Upgradeable Smart Contracts in Blockchains.".
[42] A. Zhang and X. Lin, "Towards secure and privacy-preserving data sharing in e-health systems via consortium blockchain," Journal of medical systems, vol. 42, no. 8, p. 140, 2018.

[43] Meisami, Sajad, Mohammad Beheshti-Atashgah, and Mohammad Reza Aref. "Using blockchain to achieve decentralized privacy in IoT healthcare." arXiv preprint arXiv:2109.14812 (2021).
[44] Aghabagherloo, Alireza, Javad Mohajeri, Mahmoud Salmasizadeh, and Mahmood Mohassel Feghhi. "An Efficient Anonymous Authentication Scheme Using Registration List in VANETs." In 2020 28th Iranian Conference on Electrical Engineering (ICEE), pp. 1-5. IEEE, 2020.
[45] Duan, Yue, Xin Zhao, Yu Pan, Shucheng Li, Minghao Li, Fengyuan Xu, and Mu Zhang. "Towards Automated Safety Vetting of Smart Contracts in Decentralized Applications." In Proceedings of the 2022 ACM SIGSAC Conference on Computer and Communications Security, pp. 921-935. 2022.

## AUTHORS

**Sajad Meisami** is PhD candidate in Computer Science at Illinois Institute of Technology, USA. Before that, he received his M.Sc. degree from Sharif University of Technology in Iran. His research interests include System security, Blockchain, Smart contracts and Data security and privacy.

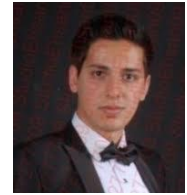

**Sadaf Meisami** is Researcher at ISSL lab at department of Electrical Engineering at Sharif University of Technology. She received her B.Sc. in Insurance Management at Management department of Kharazmi University. Her research interests include Healthcare data management and Insurance management.

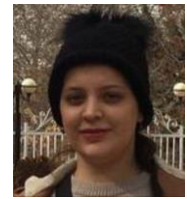

**Melina Yousefi** is Researcher at Industrial Engineering department at Isfahan University of Technology. She received her B.Sc. in Industrial Engineering department from this university. Her research interests include data management and Mathematical Finance.

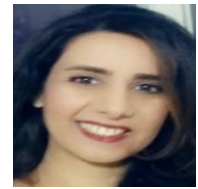

**Mohammad Reza Aref** is a professor in the department of Electrical Engineering at Sharif University of Technology. He received his PhD degree in Electrical Engineering from Stanford University, USA. His research interests include Information Theory, Cryptography theory, Statistical signal processing and Communication theory. He is a member of the IEEE.

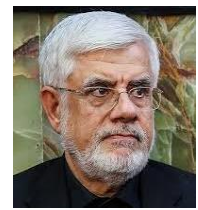